\documentclass[%
 aip,
 amsmath,amssymb,
 reprint,%
]{revtex4-1}

\usepackage{amsmath}
\usepackage{graphicx}
\usepackage{dcolumn}
\usepackage{bm}
\usepackage[mathlines]{lineno}

\usepackage{breakurl}
\usepackage[breaklinks]{hyperref}
\usepackage{CJKutf8}

\usepackage[utf8]{inputenc}
\usepackage[T1]{fontenc}
\usepackage{mathptmx}
\usepackage{etoolbox}

\makeatletter
\def\@email#1#2{%
 \endgroup
 \patchcmd{\titleblock@produce}
  {\frontmatter@RRAPformat}
  {\frontmatter@RRAPformat{\produce@RRAP{*#1\href{mailto:#2}{#2}}}\frontmatter@RRAPformat}
  {}{}
}%

\usepackage{xcolor}
\hypersetup{
    colorlinks,
    linkcolor={red!50!black},
    citecolor={blue!50!black},
    urlcolor={blue!80!black}
}

\usepackage{amsthm}
\newtheorem{theorem}{Theorem}

\def\bb{\mathbb}
\def\bf{\mathbf}

\def\R{\bb{R}}

\DeclareMathOperator*{\diag}{diag}

\makeatother
\begin{document}
\begin{CJK*}{UTF8}{gbsn}


\title[Quantum wavepackets]{Quantum wavepackets: Proofs (almost) without words}
\author{Yuxi Liu} \affiliation{Berkeley Artificial Intelligence Research Lab, University of California of Berkeley, Berkeley, California 94702}
\email{yuxi\_liu@berkeley.edu} \homepage{https://yuxi-liu-wired.github.io/}

\date{\today}

\begin{abstract}
We present a geometrical way of understanding the dynamics of wavefunctions in a free space, using the phase-space formulation of quantum mechanics. By visualizing the Wigner function, the spreading, shearing, the so-called ``negative probability flow'' of wavefunctions, and the long-time asymptotic dispersion, are intuited visually. These results are not new, but previous derivations were analytical, whereas this paper presents elementary geometric arguments that are almost ``proofs without words'', and suitable for a first course in quantum mechanics.
\end{abstract}

\maketitle
\end{CJK*}

\section{Introduction}

Phase-space quantum mechanics is a formulation of quantum mechanics that is mathematically equivalent to the standard ones (Schrödinger picture, Heisenberg picture, etc.), but offers a way of visualizing the dynamics of quantum systems that is closer to the classical intuition. In it, instead of wavefunctions that are vectors in a Hilbert space, quantum states are represented by quasiprobability distributions in phase space, called Wigner functions. 

This paper does not review phase-space quantum mechanics in detail, for which one may consult Refs \onlinecite{caseWignerFunctionsWeyl2008, curtrightConciseTreatiseQuantum2014} and references therein. To understand the paper, one only needs to know a few things about phase-space quantum mechanics.

For any pure state with position-space wavefunction $\psi(t, q)$ in $\R^n$, its corresponding Wigner quasiprobability function is defined as

\begin{equation} \label{eq:wigner}
    W(t, q, p) = \frac{1}{(2\pi )^n}\int d^n y e^{-ip \cdot y}~\psi(t, q+y\hbar/2) \psi(t, q-y\hbar/2)^*
\end{equation}

and for mixed states, the Wigner function is the linear sum of the Wigner functions of the pure states.

This allows us to calculate the probability density of finding a particle at a position in space $|\psi(t, q)|^2 = \int_\R W(t, q, p) dp$ by integration, and similarly, also its momentum-space probability density $|\psi(t, p)|^2 = \int_\R W(t, q, p) dq$.

The time-evolution is described by how $W(t, q, p)$ changes as $t$ increases, with a generalization of the classical Liouville equation:

$$
\partial_t W = \{\{H, W\}\}
$$

where $H(t, q, p)$ is a function, called the Hamiltonian of the system. Unlike the Hamiltonian operator in the Schrödinger picture, here the Hamiltonian is just a real-valued function.

If Hamiltonian $H$ is a sum of kinetic and potential energies, as $H = T(p) + V(q)$, and if both $T, V$ are quadratic polynomials, then one can show that $\partial_t W = \{H, W\}$, which is exactly the same as the classical Liouville equation of probability flow in phase space: $\partial_t \rho = \{H, \rho\}$, where $\rho$ is the (classical) probability density in phase space.

In other words, we can picture the phase-space evolution of such a Wigner function as if it is
just the classical flow of probability density in phase space, as in classical statistical mechanics. The only
difference is that there are both regions of positive and negative
"probability" densities (thus the name "quasi-probability").

There are essentially only three examples of such quadratic Hamiltonians. A particle in free space has Hamiltonian
$H = \frac{\|p\|^2}{2m}$. Thus, the Wigner function $W$ evolves by a
simple shearing flow in phase space:

$$
W(t, q, p) = W(0, q - pt/m, p)
$$

A particle under constant force $F$ has Hamiltonian $H = \frac{\|p\|^2}{2m} - F\cdot q$. Its Wigner function evolves by a parabolic translation in phase space:

$$
W(t, q, p) = W(0, q - (F t^2)/(2m), p - Ft)
$$

For example, in one dimensional space, the Airy wavepacket has a Wigner function whose contour lines are parabolas. Thus, with the right constant force, its Wigner function would remain unchanging.\cite{berryNonspreadingWavePackets1979} This makes it intuitively clear why the Airy wavepacket spontaneously accelerates, without dispersing.

Finally, a quantum harmonic oscillator in one dimension has Hamiltonian $H = \frac{p^2}{2m} + \frac 12 m\omega^2 q^2$, and so its Wigner function evolves by

$$W(t, q, p)  = W \left(0, q \cos(\omega t) - \frac{p}{m\omega} \sin(\omega t), q m \omega \sin(\omega t) + p \cos(\omega t)\right)$$

Its generalization to $n$ dimensions is immediate.

\section{Particles in free space}

\subsection{Gaussian wavepacket}

Consider the simplest case of a gaussian wavepacket on a line, centered
at $q = 0$, with zero total momentum. Over time, it contracts, until
its width $\sigma_q$ reaches a minimum, before spreading out again.
Let $t = 0$ be the time of minimal width, so its wavefunction
satisfies

$$
\psi(0, q) = \sqrt{\rho_{N (0, \sigma_q^2)}(q)}
$$

where $\rho_{N (0, \sigma_q^2)}$ denotes the probability density
function of the gaussian with mean $0$ and variance $\sigma_q^2$. By
direct calculation with Equation \ref{eq:wigner}, its Wigner function is the probability density function of the gaussian with mean $(0, 0)$ and variance
$\diag(\sigma_q^2, \sigma_p^2)$, satisfying the uncertainty principle
$\sigma_q \sigma_p = \hbar/2$.

\begin{figure*}
\centering
\includegraphics[width=\textwidth]{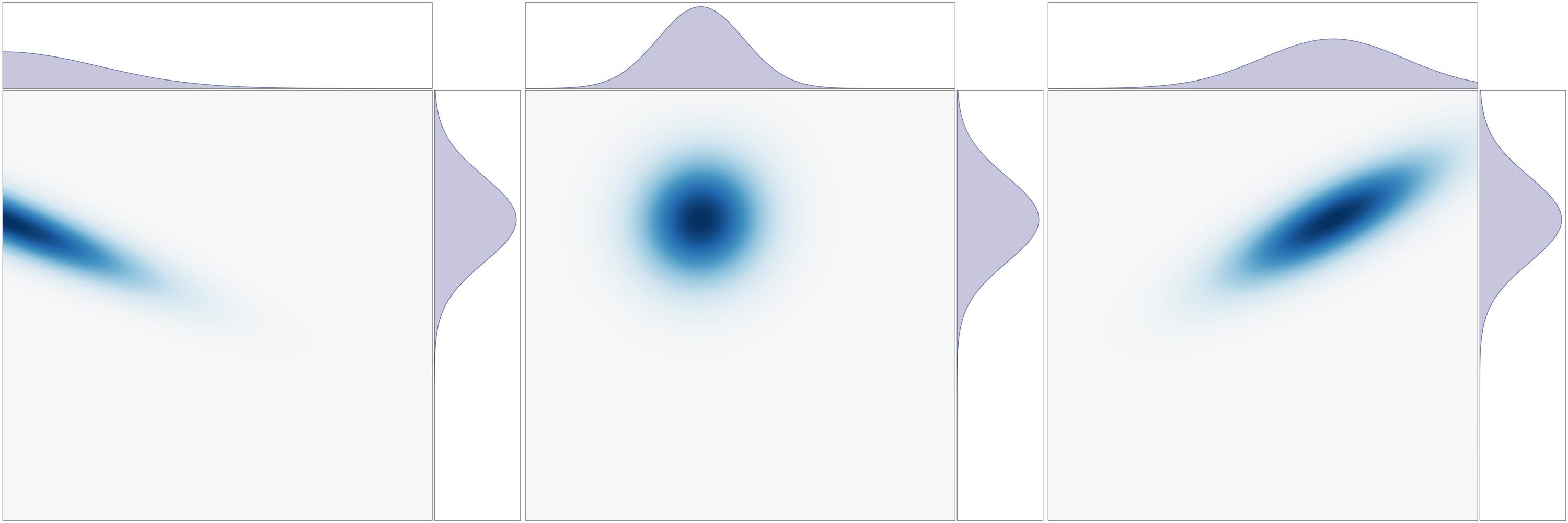}
\caption{As time increases, a gaussian
wavepacket shrinks, then spreads, on the $q$-axis marginal. Its
$p$-axis marginal remains unchanged. Colors and animation online.} \label{fig:shear_wigner}
\end{figure*}

More generally, a gaussian wavepacket with initial peak position $q_0$
and momentum $p_0$ has a Wigner function with mean $(q_0, p_0)$ and
variance $\diag(\sigma_q^2, \sigma_p^2)$. For concreteness, let
$q_0 > 0, p_0 < 0$. As time $t$ increases from $-\infty$ to
$t = 0$, the Wigner function shears to the right more and more, until
it becomes an ellipse with major axes parallel to the $p$-axis and the
$q$-axis right at $t = 0$. The center of mass on the $q$-axis is
the projection of the center of the Wigner function, which moves at
constant velocity $p_0/m$. The $q$-marginal distribution of the
Wigner function first shrinks, reaching a minimum at $t = 0$, before
growing again. Its $p$-axis marginal remains unchanged. This is shown in Figure \ref{fig:shear_wigner}.

\subsection{Negative probability flow}

Refs \onlinecite{villanuevaNegativeFlowProbability2020, goussevCommentNegativeFlow2020} noted that for a gaussian
wavepacket with positive group velocity $p_0 / m > 0$, it is often the
case that there is a paradoxical ``negative probability flow''.
Specifically, it was found that, if we stand at a point $q_1 > q_0$,
and plot $Pr(q > q_1 | t)$, the probability that the particle is found
at $q > q_1$ at time $t$, then as time passes, that probability
first decreases, before it increases, even though the wavepacket always
has positive group velocity. This is immediate in the phase space
picture.

Consider a gaussian wavepacket with
$W(0, q, p) = \rho_{N ((q_0, p_0), \diag(\sigma_q^2, \sigma_p^2))}$,
with $q_0 < 0$ and $p_0 > 0$. Geometrically, $Pr(q > q_1 | t)$ is the
integral of $W(t, q, p)$ on the region to the right of the vertical
line $q = q_1$. Instead of shearing the gaussian, we can shear the vertical line instead.
Thus, the probability is equal to the integral of $W(0, q, p)$ on
the region to the right of the sheared line $q = -\frac{t}{m} p$. The
sheared line rotates counterclockwise over time.

From the geometry of gaussian distributions, this integral can be
visually calculated by finding the contour-ellipse of
$W(0, q, p)$ that is tangent
to the sheared line. At $t \to -\infty$, the sheared line is just the
$q$-axis. As $t$ increases, the sheared line rotates
counterclockwise, and the tangent ellipse grows, until it hits the
maximal size at some extremum time $t = t_{\text{extremum}}$, at which
point the tangent ellipse is equal to

$$
\frac{(q-q_0)^2}{\sigma_q^2} + \frac{(p-p_0)^2}{\sigma_p^2} = \frac{(q_1-q_0)^2}{\sigma_q^2} + \frac{(0-p_0)^2}{\sigma_p^2}
$$

It is a simple exercise to show
$t_{\text{extremum}} = -\frac{mp_0 \sigma_q^2}{(q_1- q_0) \sigma_p^2}$.

\begin{figure*}
\centering
\includegraphics[width=0.8\textwidth]{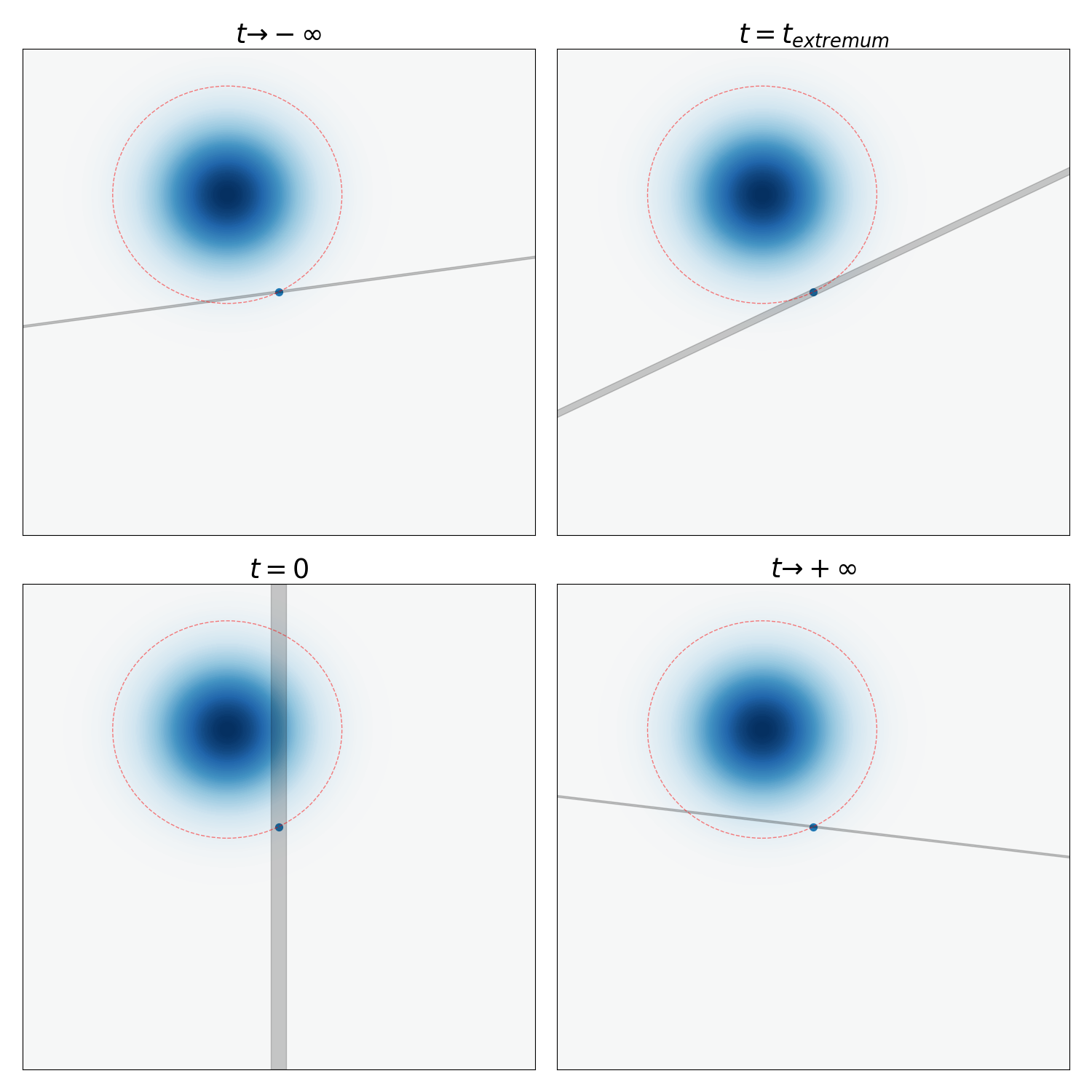}
\caption{The negative probability flow of a gaussian
wavepacket. The dashed circle is the largest possible tangent ellipse.
At $t = t_{\text{extremum}}$, the sheared line is tangent to the
dashed circle. Colors online.}
\label{fig:negative_probability_flow}
\end{figure*}

After that, the tangent ellipse shrinks again, as the sheared line rotates towards the $q$-axis again at $t \to +\infty$. This is illustrated in Figure \ref{fig:negative_probability_flow}, which makes it visually clear that $Pr(q > q_1|t)$
decreases over the region $t \leq t_{\text{extremum}}$, and increases
over the region $t \geq t_{\text{extremum}}$.

By a similar visual reasoning, if $q_0 > 0, p_0 > 0$, then
$Pr(q > q_1|t)$ increases over the region
$t \leq t_{\text{extremum}}$, and decreases over the region
$t \geq t_{\text{extremum}}$. Only when $q_0 = q_1$ is
$Pr(q > q_1|t)$ strictly monotonic over all time.

\subsection{Wave dispersion}

As noted previously, a gaussian wave packet first shrinks, then grows,
according to a precise formula proved in every introductory quantum
mechanics course. We derive this formula geometrically.

Without loss of generality, consider a wavepacket with zero group
velocity, centered at $q = 0$, reaching minimal width $\sigma_q$ at
$t=0$. Its Wigner function is
$\rho_{N((0, 0), \diag(\sigma_q^2, \sigma_p^2))}$, where
$\sigma_q \sigma_p = \hbar / 2$.

Now, consider its one-sigma ellipse
$\frac{q^2}{\sigma_q^2} + \frac{p^2}{\sigma_q^2} = 1$. It projects to
an interval $[-\sigma_q, +\sigma_q]$ on the $q$-axis, meaning that
at time $t=0$, the probability of finding the particle within the
interval $[-\sigma_q, +\sigma_q]$ is plus-or-minus one-sigma, that is,
68.3\%.

\begin{figure*}
\centering
\includegraphics[width=\textwidth]{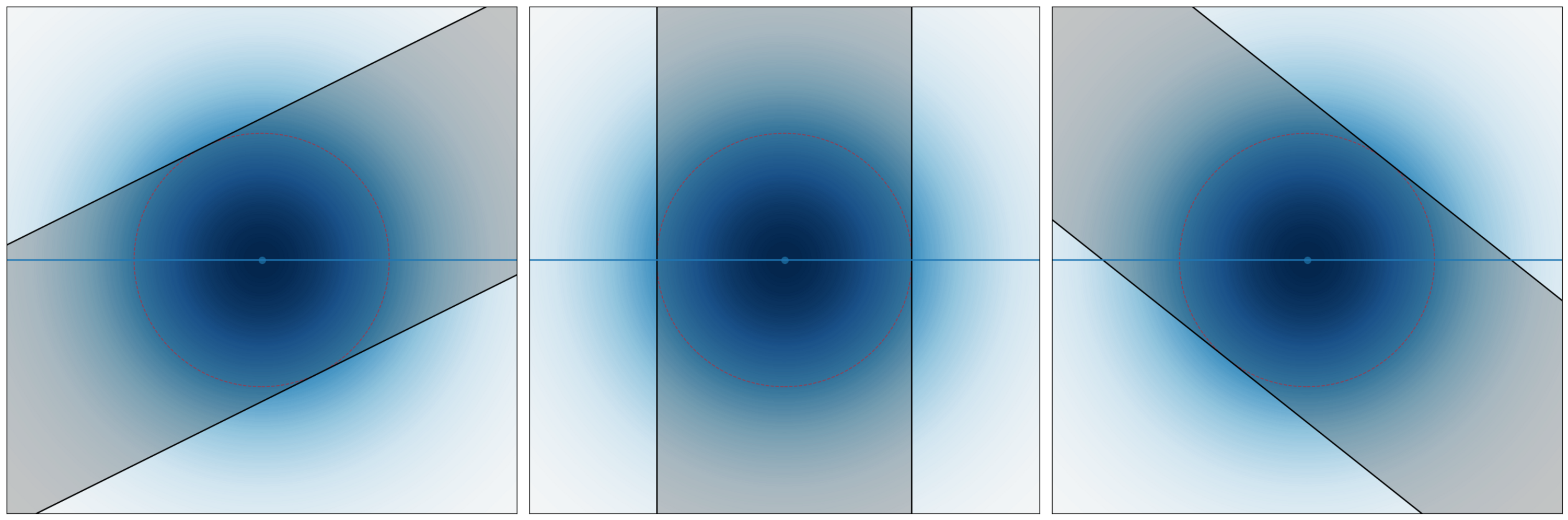}
\caption{Gaussian wavepacket spreading. As time
increases, the one-sigma circle projects to an interval on $q$-axis,
which shrinks as $t \uparrow 0$, then grows as $t \uparrow \infty$. Colors and animation online.}
\label{fig:gaussian_wavepacket_spreading}
\end{figure*}

Now, at time $t$, the new one-sigma interval can be either found by
shearing the Wigner function, or by shearing the $q$-intervals. As shown in Figure \ref{fig:gaussian_wavepacket_spreading}, the sheared $q$-intervals are the tangent lines to the
one-sigma ellipse. The tangent line has equation $q + \frac{t}{m} p = C$ for some constant $C$. By analytic geometry,\footnote{Even this minimal amount of analytic geometry can be eliminated, resulting an almost purely geometric argument. First, stretch the $q$-axis by $\sigma_q$ and $p$-axis by $\sigma_p$, turning the one-sigmal circle to a unit circle. This changes the shear line's slope angle to $-\theta$ satisfying $\tan\theta = \frac{m \sigma_q}{t \sigma_p}$. Then, the required tangent point is $(\cos\theta, \sin\theta)$. Now undo the stretching to find the tangent point $(q, p)$ in the original, un-stretched phase space.} the tangent points are

$$
(q, p) = \left( \frac{\sigma_q}{\sqrt{1 + \left( \frac{\sigma_p t}{\sigma_q m}\right)^2}}, \frac{\sigma_p}{\sqrt{1 + \left( \frac{\sigma_p t}{\sigma_q m}\right)^{-2}}}\right)
$$

and so, the projection to the $q$-axis has end points

$$
\pm \sigma_q\sqrt{1 + \left( \frac{\sigma_p t}{\sigma_q m}\right)^2} = \pm \sqrt{\sigma_q^2 + \left( \frac{\hbar t}{2m\sigma_q}\right)^2}
$$

as expected.

In particular, for very large $t$, the wavepacket spreads linearly.
This is in fact a generic result for arbitrary waves. Specifically, the
$q$-marginal of the Wigner function at $q = q_1$ and time $t$ can
be found by shearing a thin slice $q \in [q_1, q_1 + \delta q]$ back
by $\frac{pt}{m}$. If $t$ is large, then this thin slice,
perpendicular to the $q$-axis, is sheared to become almost
perpendicular to the $p$-axis instead, which is approximately
$p \in [mq_1 / t, m(q_1 + \delta q)/ t]$. Thus, we have

$$
|\psi_q(t, q)|^2 \to \frac{m}{t}|\psi_p(0, mq_1 / t)|^2
$$

as $t \to +\infty$.

\begin{figure*}
    \centering
    \includegraphics[width=1\textwidth]{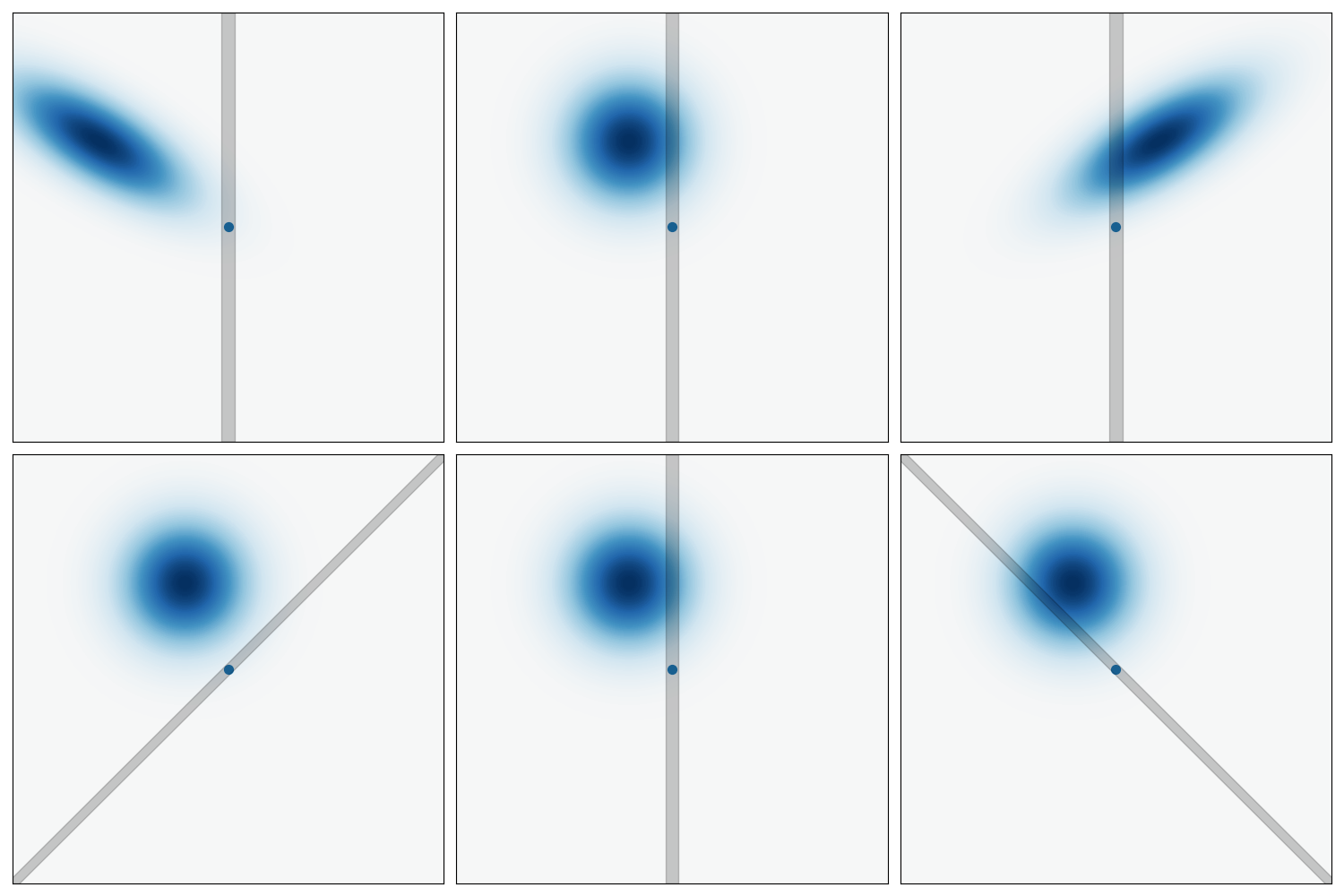}
    \caption{Shearing a gaussian Wigner function over time, then taking the marginal, is equivalent to integrating the Wigner function over vertical lines sheared in the opposite direction. Colors online.}
    \label{fig:gaussian_equivalent_shear}
\end{figure*}

This is depicted for a gaussian wavepacket in Figure \ref{fig:gaussian_equivalent_shear}.

\subsection{Hermite--Gauss waves}

For a simple harmonic oscillator with Hamiltonian
$\frac{p^2}{2m} + \frac 12 m\omega^2 q^2$, the evolution of the Wigner
function is still the same as the classical one. Thus, the Wigner
function simply rotates in phase space. A standing wave for a simple
harmonic oscillator, then, is some wavefunction $\psi(q)$ such that
its Wigner function is rotationally symmetric.

It is proved in standard introductory quantum mechanics that such standing waves are precisely the Hermite--Gauss waves:\cite[pp. 52--54]{griffithsIntroductionQuantumMechanics2018}

$$
\psi_n(x) = \frac{1}{\sqrt{2^n\,n!}}  \left(\frac{m\omega}{\pi \hbar}\right)^{1/4}  e^{
- \frac{m\omega x^2}{2 \hbar}} H_n\left(\sqrt{\frac{m\omega}{\hbar}} x \right), \qquad n = 0,1,2,\ldots.
$$

where $H_n$ are the
\href{https://en.wikipedia.org/wiki/Hermite_polynomials}{physicist's
Hermite polynomials}. Therefore, by the previous picture of how the
probability density spreads, we conclude that: The Hermite--Gauss waves are the only wavefunctions whose probability
density functions retain their shapes during propagation in free space.

Furthermore, the exact same argument as the previous section allows us
to compute how fast the wave spreads. Though the $q$-marginal is no
longer gaussian, we can still characterize its width by a single number. The precise definition does not matter, as it is only necessary to measure the shape of a Hermite--Gauss wave by a single number.

For the sake of concreteness, let $t=0$ be the point in time where the wave has minimal
spread -- that is, the point at which its Wigner function has contour
ellipses that are not tilted, but has major axes parallel to the
$q$-axis and the $p$-axis. Let $\sigma_q$ be the half-interquartile range. That is, between $q_{50\%}$ and $q_{75\%}$, where
$q_{75\%}$ is the point such that $Pr(q \leq q_{75\%}) = 75\%$. 

Since in the simple harmonic oscillator, the Wigner function just
rotates around classically. This means that we can use the classical energy conservation formula:

$$
\frac{\sigma_p(0)^2}{2m} = \frac 12 m\omega^2 \sigma_q(0)^2 \implies \sigma_p(0)/\sigma_q(0) = m\omega
$$

\begin{figure*}
\centering
\includegraphics[width=\textwidth]{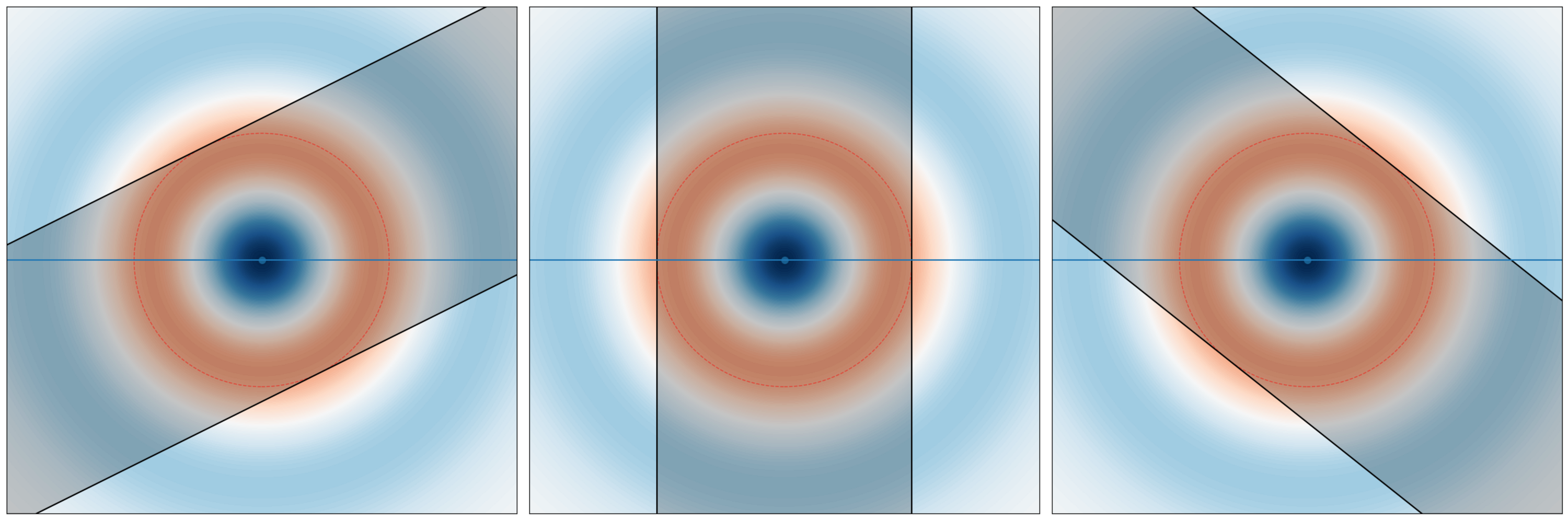}
\caption{Spreading of the $n=2$ eigenstate of the simple
harmonic oscillator. As time increases, the one-sigma circle projects to
an interval on $q$-axis, which shrinks as $t \uparrow 0$, then grows
as $t \uparrow \infty$. Colors and animation online.}
\label{fig:gaussian_wavepacket_spreading_2}
\end{figure*}

Figure \ref{fig:gaussian_wavepacket_spreading_2} shows that the same geometric argument gives

$$
\sigma_q(t) = \sigma_q(0)\sqrt{1 + \left( \frac{\sigma_p(0) t}{\sigma_q(0) m}\right)^2} = \sigma_q(0)\sqrt{1 + \left(\omega t\right)^2}
$$

which is a surprisingly simple and elegant formula. The case of the
gaussian is recovered by noting that it is the only Hermite--Gauss wave
that exactly reaches the minimum allowed by the uncertainty principle:
$\sigma_q(0) \sigma_p(0) = \hbar/2$, which, when combined with
$\sigma_p(0)/\sigma_q(0) = m\omega$ that is satisfied by all
Hermite--Gauss waves, gives us
$\omega = \frac{\hbar}{2 m \sigma_q^2}$, and so we recover the
previous result of
$\sigma_q(t) = \sigma_q(0)\sqrt{1 + \left(\frac{\hbar t}{2 m \sigma_q^2}\right)^2}$

Ref \onlinecite{andrewsEvolutionFreeWave2008} proved these results analytically. However, as far as the
author knows, this is the first time these were derived geometrically, with minimal calculus.

\subsection{Square wave}

As a concrete example of a wave that is not Hermite--Gauss, consider the square wave function

$$
\psi(q) = \begin{cases}a^{-1/2} & \text{ if }q \in [-a/2, +a/2]\\ 0  & \text{ else}\end{cases}
$$

Its Wigner function is easily calculated as

$$
\begin{aligned}
      W(0, q, p) &= \frac{1}{2a\pi \hbar} \int_{|q + y/2| \leq a/2, |q - y/2| \leq a/2} dy e^{-ipy /\hbar} \\
      &= \frac{1}{2a\pi \hbar} \frac{\hbar}{-ip} e^{-ipy /\hbar}\Big|_{\max(2q - a, -2q- a)}^{\min(2q+a, -2q+a)}\\
      &= \frac{1}{a \pi p} \sin\left(\frac{p(a - 2|q|)}{\hbar} \right)
      \end{aligned}
$$

when $q \in [-a/2, +a/2]$. For larger values, $W = 0$.

The $p$-marginal distribution is

$$
\begin{aligned}
\rho_p(p) &:= \int_\R  W(0, q,p)dq  \\
      &= \int_{-a/2}^{+a/2} \frac{1}{a \pi p} \sin\left(\frac{p(a - 2|q|)}{\hbar} \right) dq \\
      &= \frac{\hbar }{\pi a p^2} ( 1 - \cos(pa /\hbar))
\end{aligned}
$$

\begin{figure*}
\centering
\includegraphics[width=0.8\textwidth]{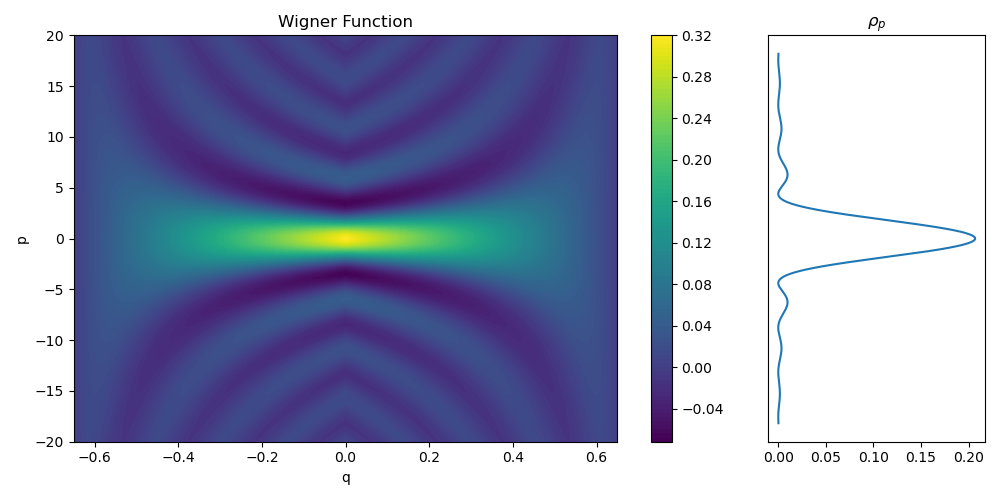}
\caption{The Wigner function of the square wave and its $p$-marginal
distribution. $a = 1.3, \hbar = 1$. Colors online.} \label{fig:square_wave_wigner}
\end{figure*}

The Wigner function and its $p$-marginal are shown in Figure \ref{fig:square_wave_wigner}.

Let
$f(t) := \frac{1 - \cos t}{\pi t^2} = \left(\frac{\operatorname{sinc}(t/2)}{\sqrt{2\pi}} \right)^2$,
then $\rho_p(p) = \frac{a}{\hbar} f(\frac{a}{\hbar}p)$. Thus, the
$p$-marginal distribution always has the same shape no matter what
$a$ is.

Because at $t=0$, the Wigner function is zero outside of
$q \in [-a/2, +a/2]$, at large $t$, the probability that the
particle is found at $[q, q + \delta q]$ is the integral of $W(0, \cdot, \cdot)$
over a the thin line passing $[q, q + \delta q]$ at slope $-m/t$.
Thus, we have

$$
\begin{aligned}
\int_\R W(t, q, p)dp 
    &\approx \frac mt\int_{-a/2}^{+a/2} W(0, q, p)dq \\
    &= \frac{m}{t} \rho_p(qm/t) = \frac{\hbar t}{\pi m a q^2} \left(1 - \cos \left( \frac{ma}{\hbar t} q\right)\right) \\
    &= \frac{am}{\hbar t} f \left(\frac{am}{\hbar t} q\right)
\end{aligned}
$$

In particular, the higher-order waves are \emph{not} disspated away, and
so the square wave never converges to a gaussian wavepacket. In fact, it
always converges to the same shape of $f$. This is directly against
the claim of Ref \onlinecite{mitaDispersionNonGaussianFree2007}, as previously
pointed out by Ref \onlinecite{andrewsEvolutionFreeWave2008}.

By the symmetry of the Wigner function in $q$ and $p$, we have the
following result: If the wavefunction satisfies
$\psi(q) = \sqrt{\frac{b}{\hbar} }\frac{\operatorname{sinc}(bq/2 \hbar)}{\sqrt{2\pi}}$
at $t=0$, then at large $t$, the wavefunction converges to a square
wave on the interval $[-bt/2m, +bt/2m]$ with height
$\sqrt{\frac{m}{bt}}$.

\subsection{Generic wavepacket}

In general, the shape of the wave of any normalizable wave function
$\psi_q(q)$, propagating in free space, would, after a long time,
converge to a translation-and-dilation of its $p$-marginal
distribution $\psi_q(q)$, multiplied by a phase factor
$e^{iS(t, q)}$. Since the $q$-marginal distribution can be
arbitrary, the same is true for the $p$-marginal distribution. In
particular, we have the following theorem, which is immediate from the
geometric intuition.

\begin{theorem}

If $\rho$ is a smooth probability density function on $\R$, then
there exists a wavefunction $\psi$ propagating in free space, such
that for all large enough $t$,

$$
|\psi(t, q)|^2 \to \frac{m}{t}\rho(mq/t)
$$
\end{theorem}

\begin{proof}

Such $\psi(0, q)$ can be found by taking the Wigner function of
$\sqrt{\rho(p)} e^{iS(p)}$ for an arbitrary smooth function $S$,
rotating by 90 degrees in the phase plane, then taking the inverse
Wigner transform.

\end{proof}

Previously, we showed that for almost any gaussian wavepacket,
$Pr(q > q_1 | t)$, i.e. the probability of finding the particle to the right
of some point $q_1$ at time $t$, would first increase then decrease,
or first decrease then increase. In fact, this is true for almost any
Wigner function, period. The Wigner function does not even need to be
pure. It can be the Wigner function of a mixed state, and this would
still be true.

\begin{theorem}

For almost any mixed state of a particle propagating in free space, and
almost any $q_1$, the function $t \mapsto Pr(q > q_1 | t)$ is not
monotonic.

\end{theorem}

\begin{proof}

Fix some Wigner function $W$. Define $f(\theta)$ to be the integral
of $W$ to the right of the line passing $(q_1, 0)$, making an angle
$\theta$ with the $q$-axis. Then we have $f(0) = 1 - f(\pi)$, and
in general, $f(\theta) = 1 - f(\pi + \theta)$. Thus, if $f$ reaches
a global minimum at $\theta$, then it reaches a global maximum at
$f(\theta + \pi)$. In particular, if $f(0)$ is neither the global
minimum nor the global maximum, then $f(\theta)$ must reach either the
global minimum or the global maximum at some $\theta \in (0, \pi)$.
Therefore, $f$ is not monotonic as $\theta$ rotates from $0$ to $\pi$.

\end{proof}

We can think of this as a ``maximally non-dissipation'' result. Unlike
the heat equation, where all high-frequency fluctuations decay away,
leaving behind just a single low-frequency gaussian mode, the
Schrödinger equation results in a wave propagation that has no
dissipation, but preserves the shape of $|\psi_p(0)|^2$ for all large
times. This is unsurprising, as it is a common folklore that quantum mechanical wavefunctions, like classical waves, do not dissipate.

\section{Conclusion}

The dynamics of wavefunctions in free space, while often covered in introductory courses using analytical tools, can be somewhat unintuitive. Switching to the phase-space formulation of quantum mechanics, while not traditionally done, offers a way of visualizing these dynamics using the Wigner function. The spreading, shearing, the so-called ``negative probability flow'' of wavefunctions, and the long-time asymptotic dispersion of waves, become visually intuitive. Educators may find it pedagogically useful to visualize the Wigner function, alongside the more traditional analytical derivations, in an introductory quantum mechanics course. The material is perhaps even teachable to gifted high-school students.

\begin{acknowledgments}

Some plotting code was modified from a description in the Wikimedia Commons.\cite{nebulaEnglishWignerQuasiprobability2024} Descriptions on Wikimedia Commons are available under CC0 license.

\end{acknowledgments}

\section*{Data Availability Statement}

The plotting code and animations are available both as supplementary material available at the official website American Journal of Physics and at the author's personal website.

\nocite{*}
\bibliography{AIP_gaussian_wavepacket}

\end{document}